\documentclass[11pt, manuscript, nonacm]{acmart}

\usepackage[english]{babel}
\theoremstyle{plain}

\theoremstyle{definition}

\theoremstyle{remark}

\usepackage{xcolor}

\definecolor{MyBlue}{RGB}{0, 64, 64}

\AtBeginDocument{%
	\providecommand\BibTeX{{%
			\normalfont B\kern-0.5em{\scshape i\kern-0.25em b}\kern-0.8em\TeX}}}
\begin{document}
	\title{Consensus In Asynchrony}
	
	\author{Ivan Klianev}
	\email{Ivan.Klianev@gmail.com}
	\affiliation{%
		\institution{Transactum Pty Ltd, Sydney NSW}
		\country{Australia}
	}
	
	\title{\textcolor{MyBlue}{Consensus In Asynchrony}}
	
	
	\begin{abstract}
	We demonstrate sufficiency of events-based synchronisation for solving deterministic fault-tolerant consensus in asynchrony. Main result is an algorithm that terminates with valid vector agreement, hence operates with safety, liveness, and tolerance to one crash. Reconciling with the FLP impossibility result, we identified: i) existence of two types of agreements: data-independent and data-dependent; and ii) dependence of FLP theorem correctness on three implicit assumptions. Consensus impossibility with data-dependent agreement is contingent on two of them. The theorem-stated impossibility with every agreement type hinges entirely on the third. We provide experimental results showing that the third assumption has no evidence in support.	
	\end{abstract}

	
	
	\maketitle
	
	\textcolor{MyBlue}{\section{INTRODUCTION}}
	
		The work of Fischer, Lynch, and Paterson (FLP) \cite{FLP_result} discovered existence of a problem with fault-tolerant distributed computing in total asynchrony. The possibility of a crash may lead to a bivalent configuration where binary consensus seems impossible. The paper demonstrated the inevitability of existence of such configuration and stated it as evidence proving the impossibility of fault-tolerant consensus in asynchrony.
		
		Forty years later, the exploration for the possibility of a fault-tolerant deterministic consensus in asynchrony is still considered a direction with a dead end. Yet the works on linearisability \cite{Herlihy_Wing_1990}, atomic consistency and state machine replication \cite{Lamport_StateMachine} \cite{Lampson_1996} give a hint. When processes compute their decision values from an atomically consistent state, they terminate with the same agreement value. Hence, consensus about the initial input values of the system processes is the key to all types of consensus.
		
		We solved consensus about the initial input values with a crash-tolerant algorithm that terminates in total asynchrony with vector agreement \cite{PeaseShostakLamport80}. It neutralises the effects of asynchrony with events-based synchronisation and terminates binary consensus with agreement about the initial inputs. With a binary value assigned to \textit{tie state}, the processes individually compute the binary agreement value regardless of the content of the agreed vector of initial inputs.

	\textcolor{MyBlue}{\subsection{\textsf{Background}}}
		
		Fischer, Lynch, and Paterson demonstrated non-termination in the following way: i) Set out that a consensus protocol is \textit{partially correct} if it terminates \textit{some} runs with one decision value, and it is \textit{totally correct} if \textit{every} run is partially correct; ii) Proved that a protocol starting with randomly assigned 1-bit values and aiming a binary decision has at least one \textit{bivalent} initial configuration of its starting data, i.e., may reach a configuration where one missing message can change the consensus outcome; iii) Presented a scenario where an unspecified consensus protocol is not totally correct in asynchrony in spite of one crash. In this scenario, the correct processes wait infinitely long for a missing message critical for decision-state, being unaware that what causes its missing is not asynchrony but inability of a crashed process to send it. The paper declares that if “consensus problem involves an asynchronous system of processes”, then “\textit{every} protocol for this problem has the possibility of non-termination, even with only one faulty process”.
		
	\textcolor{MyBlue}{\subsection{\textsf{Motivation}}}
	
		Binary consensus is critical for atomicity of third-party coordinated distributed transactions \cite{JimGray78} \cite{MohanStrongFink1985} \cite{HectorGM_EtAl1986}. This coordination relies on a choice from a family of distributed commit protocols \cite{JimGray78} \cite{BruceLindsay79} \cite{LampsonSturgis76}, each susceptible to blocking \cite{BernsteinHadzilacosGoodman}. Binary consensus non-termination is one of the reasons for blocking \cite{Gray_lamport}.
	
		A decentralised database system operates with global replication of local transactions and ensures atomic consistency across local systems with state machine replication \cite{Lamport_StateMachine} \cite{Schneider_1990} \cite{Lampson_1996}. Agreement for its purposes is about a vector containing state. Agreeing about a vector is also a critical element in our handling of asynchrony.
	
		Proving termination of binary consensus in complete asynchrony with tolerance to one crash would naturally challenge the axiomatic acceptance of the influential, both in theory and practice, conjecture that no consensus protocol can operate with ensured simultaneous safety, liveness, and fault-tolerance in partial synchrony \cite{GilbertLynch_2012}.
		
	\textcolor{MyBlue}{\subsection{\textsf{Highlights}}}
	
		This paper:
	
	\begin{itemize}
	
		\item Points out what makes the FLP theorem formally correct – its own set of implicit assumptions.
	
		\item Identifies the existence of two types of consensus agreements – data-dependent, e.g., binary agreement, and data-independent, e.g., vector agreement.
	
		\item Presents a deterministic crash-tolerant consensus algorithm that operates with vector agreement in total asynchrony.
	
		\item Demonstrates that the algorithm terminates with valid vector agreement, i.e., agreement about the initial inputs of the processes.
	
		\item Demonstrates that a crash-tolerant consensus with binary agreement can terminate in asynchrony with vector value and then compute the binary value.
	
		\item Points out that the FLP impossibility result with binary agreement critically depends on assumption that \textit{tie score} indicates impossible binary agreement. 
		
		\item Points out that the FLP impossibility to terminate consensus aiming binary agreement critically depends on assumption that a binary consensus protocol cannot terminate in asynchrony before deciding on the binary agreement value.
	
		\item Points out that the FLP impossibility of \textit{any} fault-tolerant consensus hinges entirely on an unchecked assumption that the impossibility of data-dependent type of consensus demonstrates also impossibility of the data-independent type.
	
	\end{itemize}		
	
	\textcolor{MyBlue}{\subsection{\textsf{Problems Formulation}}}
	
		In a totally asynchronous environment and a consensus system of $n$ processes, $n\ge5$, tolerating possible crash of one process, and every process $P_i$, $i\in\{1,...,n\}$, has an initial value $v_i\in\{0,1\}$, demonstrate that no less than $ (n-1) $ processes can agree on an $n$-vector $ V $, where $V=(v_1,...,v_n)$, with at most one element of vector $ V $ replaced with a null marker $ \varnothing $, denoting an empty element.
	
		Demonstrate also that: i) with an assigned binary value to \textit{tie state} a consensus with binary agreement reduces to consensus with vector agreement; and ii) the impossibility of binary consensus is caused entirely by not having a binary value assigned to the \textit{tie state} and thus not allowing it to be a legitimate agreement value.
	
	\textcolor{MyBlue}{\subsection{\textsf{Applied Methods}}}
	
		Our work aimed to demonstrate the possibility for deterministic crash-tolerant consensus in total asynchrony. In view of the decades-long period of accepting that complete impossibility of any fault-tolerant consensus in asynchrony is thoroughly proven as unconditional, we were fully aware that this is not sufficient. The results must be reconciled with the factual evidence in the FLP theorem proof. 
		
		In pursuit of these objectives we applied a variety of methods. Critical analysis of the FLP theorem proof revealed its dependence on unstated but implied assumptions, some unchecked, that make the theorem formally correct. It also revealed that the theorem is presented as a hypothesis with conditional supporting evidence. We challenged this hypothesis with a counter-hypothesis having stronger evidence of support.
		
		Research through design was the only method allowing to demonstrate that the events-based synchronisation is sufficient for the purposes of deterministic consensus in total asynchrony. It allowed to analytically explore for existence of a possibility to terminate a crash-tolerant vector consensus algorithm with a valid agreement about the initial input values of system processes, by using only events-based synchronisation.
		
		We also designed and built a synchronous consensus system, and used it to experimentally collect data allowing us to explore for existence of correlation or causality between vector and binary agreements. Moreover, to explore whether the demonstrated by the FLP theorem impossibility of binary agreement is caused by the total asynchrony or by the interpretation of \textit{tie state} as "undecidable".
		
		For reconciliation purposes we contrasted what the proof of the FLP theorem in fact demonstrates and what must be demonstrated in order to prove impossibility of fault-tolerant consensus with vector agreement about the inputs from the system processes. Thus we analysed whether the logical contradiction between the FLP impossibility and the possibility we demonstrate indicates that one of these results must be wrong.
		
		At the end, we produced a conclusion based on the entire set of factual and analytical evidence collected with these methods. It summarises whether the nature-established limits allow fault-tolerant consensus in total asynchrony possible or not.
		
	\textcolor{MyBlue}{\section{CRITICAL ANALYSIS}}
		
		FLP theorem states: In total asynchrony “No consensus protocol is totally correct in spite of one fault”. In view of the specific results demonstrated in the theorem's proof, this claim is unreasonably strong. It hinges on three unstated assumptions: the \textit{impossibility to agree}, the \textit{impossibility to terminate}, and the \textit{equal impossibility} of binary consensus and vector consensus. Moreover, its proof is credible only within a tiny area that satisfies five restrictions.
		
	\textcolor{MyBlue}{\subsection{\textsf{Assumption: Impossibility to Agree}}}
	
		Using a rule of arithmetic, it is impossible to always produce a binary result from a random set of data. Counting the votes of an even number of voters for two candidates occasionally scores a tie. Regardless of how small the probability is for a tie, its possibility is certain. So, \textit{tie score} does not indicate conceptual impossibility of consensus due to impossibility of agreement. Needed is just to assign a binary value to it.
		
		A binary consensus protocol, with no binary interpretation of \textit{tie state}, cannot ensure crash-tolerant agreement with an odd number of processes and cannot ensure agreement with an even number of processes with no crash. Such a protocol cannot prevent the "undecidable" data configurations with these two system configurations.
		
		\textbf{Impossibility to agree} assumption: A \underline{tie score} result produced by binary consensus protocol, which computes the outcome by counting the 0s and the 1s, \underline{indicates} impossibility to agree on a binary outcome value. 
		
		No reason has ever been presented about why tie score is considered impossibility of agreement, having in mind that it is an entirely legitimate outcome. Moreover, with this assumption a system of an odd number of processes cannot tolerate a fault – impossibility to agree is unavoidable with input from an even number of processes.
	
	\textcolor{MyBlue}{\subsection{\textsf{Assumption: Impossibility to Terminate}}}
	
		This assumption follows from the impossibility to agree on a binary outcome value. Termination in asynchrony cannot happen before deciding on the agreement value.
		
		\textbf{Impossibility to terminate} assumption: A binary consensus protocol \underline{cannot} terminate in asynchrony \underline{before} deciding on the \underline{binary} agreement value.
	
		Later in the paper we demonstrate why this assumption is incorrect. We present a crash-tolerant consensus algorithm, show its termination in total asynchrony with vector agreement, and show that the processes in a binary consensus system can terminate with vector agreement and then compute binary agreement.
		
		The processes, having terminated with agreement on a vector of initial input values, can use the content of this vector to individually compute the binary agreement value, confident that they can't go wrong for the following reason. Starting from atomically consistent state \cite{Herlihy_Wing_1990} and computing with the same deterministic procedure, the processes individually produce \cite{Lamport_StateMachine} \cite{Schneider_1990} \cite{Lampson_1996} atomically consistent state.
		
	\textcolor{MyBlue}{\subsection{\textsf{Assumption: Equal Impossibility of Consensus}}}

		The theory of distributed computing has not yet acknowledged the existence of two different consensus types: data-dependant and data-independent. Here is a brief introduction of the concept.

		\textbf{Definition 1}: \textit{Data-Dependant Consensus}. A type of consensus where the possibility for agreement depends on the content of the initial input of the system processes and the faults that may affect inclusion of every initial input in the decision making state.

		\textbf{Definition 2}: \textit{Data-Independent Consensus}. A type of consensus where the possibility for agreement is not affected by the actual content of the initial input of the system processes.

		A data-independent consensus protocol executes steps necessary for a required number of processes to reach agreement about a set of data. A data-dependent consensus protocol does the same and agrees in some form on the same set of data. Yet, the agreement still may not be achieved due to the content of the agreed data. The data-dependent type is exemplified by protocols operating with binary agreement. The data-independent type is exemplified by protocols operating with vector agreement.

		\textbf{Equal impossibility} assumption: An impossibility result demonstrated with data-dependent consensus is applicable to data-independent consensus protocols. 
		
		Existence of any grounds for such an assumption has never been demonstrated. Later in the paper we present experimental results showing that it is incorrect. A synchronous consensus system was created for the purpose of exploring the relationship between binary and vector agreements. Analysis of the experimental data, collected with this system, shows why this assumption is not correct.
		
	\textcolor{MyBlue}{\subsection{\textsf{New Perspective on the Impossibility Results}}}
	
		The two major consensus impossibility results in distributed computing, FLP in total asynchrony and \cite{SantoroEtAl1989} in complete synchrony,
		were proven under the \textit{impossibility to agree} assumption. Yet both were claimed as valid beyond binary agreement under the \textit{equal impossibility of consensus} assumption.
		
		It is a prerogative of the authors of both works to decide whether or not to make the assumption that \textit{tie state} reflects impossibility of binary consensus. Both works argue for impossibility of consensus protocol to terminate with valid agreement by pointing out impossibility to decide on the value of binary agreement. The reason behind this is hard to reject or accept. Yet the assumption about equal impossibility of consensus has no reasonable grounds. To assume that impossibility valid with data-dependent consensus is also valid with data-independent consensus protocols requires presenting some evidence in support.
		
		It is easy to examine a conjunction of the two assumptions for correctness. To prove that at least one of the two is incorrect, it is sufficient to point out that the conjunction of both can produce an obvious or provably wrong result. With the above two assumptions it can be "proven" that consensus in complete synchrony is impossible in complete absence of any faults, which is clearly nonsense.
		
	\textcolor{MyBlue}{\subsection{\textsf{Conditionality of the FLP Proof}}}
	
		Without the \textit{impossibility to agree} assumption, the FLP impossibility cannot be proved. With only the \textit{impossibility of agree} assumption, the FLP impossibility is valid inside a tiny area of the entire space of possibilities as its proof hinges on \textbf{five} restrictions. The rest of the paper shows that the impossibility result is not applicable where one of these two assumptions is not presented.  
		
		Here is list of the restrictions making the proof of FLP theorem credible:
		
		\textbf{Restriction 1}. The term \textit{consensus protocol} comprises only those that operate with a data-dependent agreement.
		
		\textbf{Restriction 2}. The data-dependent consensus protocols under Restriction 1 are only those with a binary agreement having no binary value assigned to the \textit{tie state}.
		
		\textbf{Restriction 3}. The binary agreement protocols under Restriction 2 are only those that do not terminate with agreement about the state necessary for computing the binary agreement value.
		
		\textbf{Restriction 4}. The fault-tolerant protocols under Restriction 3 are only those that ignore the required necessity to be able to terminate with crash of one process.
		
		\textbf{Restriction 5}. The protocols, that ignore the necessity to be able to terminate with a crash under Restriction 4, are only those that do not use any of the existing possibilities to terminate.
		
		The only substantial evidence in the proof is the demonstrated inevitable existence of the so-called \textit{bivalent} configuration. In fact this configuration is \textit{trivalent} for any fault-tolerant consensus protocol, in asynchrony or not. A protocol designed to tolerate one crash must be prepared to terminate with \textit{tie state}.
		
	\textcolor{MyBlue}{\section{THE TWO HYPOTHESES}}
	
	
	\textcolor{MyBlue}{\subsection{\textsf{The Hypothesis of Impossibility}}}
	
		FLP theorem demonstrates inability of an \textit{unspecified} consensus protocol to guarantee termination with binary agreement in total asynchrony with tolerance to one crash. Yet the theorem claims that its proof demonstrates inability of \textit{every} crash-tolerant consensus to terminate in asynchrony. Even more, it does not provide any reason why, as if the \textit{equal impossibility of consensus} is obvious to everyone. 
	
		We respectfully disagree that the proof contains any evidence 
		confirming, or at least supporting, this hypothesis for the following reasons:
	
		\textbf{Reason 1}: The FLP paper presents a result shown with data-dependent consensus as valid with both consensus types. Thus it claims that something made impossible by a set of restrictions is also impossible where these restrictions do not exist.	
	
		\textbf{Reason 2}: The theorem's protocol is inapplicable for the purposes of fault tolerance in asynchrony. It broadcasts initial input, waits to collect $n$ inputs, and blocks on coincidence of a bivalent configuration and a crash. If it were made to tolerate a crash, it would have checked for the possibility to terminate with inputs from $(n-1)$ processes instead of waiting.
		 
		Here is how. The interactive consistency problem, where all correct processes must agree on a vector of values proposed by all processes, is considered harder \cite{Reynal_2005} than the consensus. Yet it can be solved in the FLP model with simple implementation of failure detection based on analysis of the content of the received messages. The model assumes "atomic broadcast", where "a process can send the same message in one step to all other processes with the knowledge that if any non-faulty process receives the message, then all the non-faulty processes will". This assumption is sufficient to demonstrate that the blocking is avoidable. Every process having received $(n-2)$ initial inputs atomically broadcasts an $n$-vector $V=(v_1,...,v_n)$, where the missing value $v_x$ is replaced with a null marker $ \varnothing $. A received vector with $ \varnothing $ on a different position indicates that all processes did broadcast initial input. Alternatively, $(n-2)$ received vectors with $ \varnothing $ on the same position lets the protocol terminate, regardless of what causes the missing message. Later in the paper we show how to ensure that all healthy processes decide the same vector, including the process whose initial input value is not included as an element in the decided vector.
	
		\textbf{Reason 3}: Even if the theorem's protocol were applicable for its purpose, the eventual proof with it would not have confirmed impossibility. Proving a problem is impossible to solve requires analysis of \textit{all} approaches promising to solve it. Confirming the hypothesis of impossibility, requires analysis of \textit{all} known and unknown alternatives of the used protocol. Showing impossibility with each one, however, is not feasible without knowing how many are the \textit{unknown}, yet possibly \textit{existing} under the laws of nature, alternatives. The hypothesis is unconfirmed in this aspect.
	
		Later in the paper we present one such alternative. It compels us to question the strong claim of the theorem. We respectfully disagree that fault-tolerant consensus with binary agreement, and even less fault-tolerant consensus with vector agreement, has no theoretical solution in asynchrony.

	\textcolor{MyBlue}{\subsection{\textsf{The Counter-Hypothesis}}}
	
		\textbf{Counter-Hypothesis}: Deterministic crash-tolerant consensus algorithm can terminate with valid vector agreement in total asynchrony and thus simultaneously ensure each one of the following operational properties: 
	
	\begin{itemize}
		\item Safety: Never produces invalid agreement value;
		\item Liveness: Always terminates; and
		\item Crash-Tolerance: One crash cannot compromise safety or liveness.
	\end{itemize}
	
		\textbf{Definition 3}: \textit{Vector Agreement.} Agreement where $ n $ processes $P_i$, $i\in\{1,...,n\}$: a) start with individual initial value $v_i$ that can be a vector with finite number of elements; b) individually compose an $ n $-vector $V_i=(v_1,...,v_n)$; c) terminate when $\ge (n-f)$ processes $P_i$ agree on a vector $V=(v_1,...,v_n)$ containing at most $f$ empty vector elements; and d) satisfies the following properties:
	
	
	\begin{itemize}
		\item Agreement: All correct processes agree on the same vector;
		\item Validity: The agreed-upon vector contains $\ge (n-f)$ initial values as vector elements;
		\item Termination: Every correct process reaches a decision.
	\end{itemize}
	
		The evidence confirming our hypothesis is demonstrated in the following manner. After defining the system model, we present our algorithm which establishes agreement, as per Definition 3, with tolerance to one unannounced crash. The proof of its termination is the confirming evidence. 
	
	\textcolor{MyBlue}{\subsection{\textsf{System Model}}}
	
	A consensus protocol operates in a totally asynchronous system of $ N $ processes $(N \geq 5)$. The protocol cannot make any assumptions about relative speed of processes, about delay time in delivering a message, or about order of delivery of messages.
	
	The processes communicate with \textit{atomic broadcast}. So, a process can send the same message in one step to all other processes with the knowledge that if any non-faulty process receives the message, then all the non-faulty processes will receive it.
	
	Each process $P$ has an \textit{input register} $R^{In}$, an \textit{output register} $R^{Out}$, and an unbounded amount of internal storage. Register $R^{Out}$ is a vector of $N$ elements $R^{In}$: $R^{Out}=( R^{In}_{1},R^{In}_{2},...,R^{In}_{N} )$.
	
	\textit{Internal state} of a process $ P $ comprises value in $R^{In}$ and value in $R^{Out}$, together with a program counter and internal storage.
	
	\textit{Starting values} of the internal state are fixed, with an exception for the starting value in $R^{In}$. More specifically, $ R^{In} $ starts with a \textit{length prefixed string} having an arbitrary value from a finite universe $ S $, and $ R^{Out} $ starts always with the fixed value	$R^{Out}=( \varnothing,\varnothing,...,\varnothing )$, where $ \varnothing $ is a null marker denoting an empty element.
	
	\textit{Objective} is to establish agreement between $\geq (N-1)$ correct processes about the content of their individual output registries $ R^{Out} $ in the possibility of one crash.
	
	\textit{Proposal} is a vector of $ N $ elements that contains at most 1 element with value $\varnothing $. 
	
	\textit{Decision} is also a vector of $ N $ elements that contains at most 1 element with value $\varnothing $. 
	
	\textit{Decision state} is an internal state a process $ P $ reaches after it has broadcast a decision and has collected $ (N-1) $ equal decisions. Once a decision state is reached, $P$ writes its value in register $ R^{Out} $.
	
	\textit{Decision value} is the actual vector value which $P$ writes in $R^{Out}$ after reaching a decision state.
	
	$ P $ acts deterministically according to a transition function, which cannot change the value of $R^{Out}$ once $ P $ reaches a decision state. The system is specified by the transition function $ \Phi $ associated with each of the processes and the initial values in the input registers. The processes strictly follow function $ \Phi $, i.e., there is no Byzantine behaviour.
	
	The system operates with no faulty links. Processes communicate by sending each other messages. A \textit{message} is a tuple $(P_{s}, P_{d}, m)$, where $P_{s}$ is the sender process,  $ P_{d} $ is the destination process, and $ m $ is a 'message value' from a fixed universe $ M $. The message system maintains a \textit{message buffer} – a multi-set of messages that have been sent but not yet delivered. It supports two abstract operations:
	
	- \textit{Send}($P_{s}, P_{d}, m$). It places ($P_{s}, P_{d}, m$) in the message buffer.
	
	- \textit{Receive}($P_{d}$). It deletes some message ($P_{s}, P_{d}, m$) from the buffer and returns $ m $, in which case ($P_{s}, P_{d}, m$) is \textit{delivered}, or $ \varnothing $ and leaves the buffer unchanged.
	
	Thus, the message system acts non-deterministically, subject only to the condition that if \textit{Receive}($P_{d}$) is performed infinitely many times, then every message ($P_{s}, P_{d}, m$) in the message buffer is delivered. In particular, the message system is allowed to return $ \varnothing $ a finite number of times in response to \textit{Receive}($P_{d}$), even though a message ($P_{s}, P_{d}, m$) is present in the buffer.
	
	Basic terms:
	
	- \textit{Configuration}. A set containing the internal state of every system process and the contents of its message buffer.
	
	- \textit{Step}. A step takes one configuration to another and consists of a primitive step by a single process $ P $. A step executes a sequence of activities as one atomic operation. Let $ C $ be a configuration. A step occurs in two phases. First, \textit{Receive}($P$) on the message buffer in $ C $ obtains a value $m \in M \cup |\varnothing|$. Next, depending on the internal state of $ P $ in $ C $ and on value $ m $, process $ P $ may enter a new internal state and may send a finite set of messages to other processes. Since the processes are deterministic, a step of $ P $ is completely determined by the internal state of $ P $ and the received value $ m $.
	
	- \textit{Event}. A tuple ($P_{s}, P_{d}, m$) indicating delivered $ m $ from $P_{s}$ to $P_{d}$.
	
	- \textit{Schedule}. A schedule from configuration $ C $ is a finite or infinite sequence of events that can be applied in turns, starting from $ C $. For a process $ P $ schedules are commutative \cite{FLP_result}, until $ P $ sends a proposal.
	
	- \textit{Run}. A sequence of steps taken in association with a schedule.
	
	- \textit{Correct process}. A process is correct in a run if it takes infinitely many steps. Otherwise it is \textit{faulty}.
	
	Configurations:
	
	- \textit{Initial}. A configuration in which each process starts at the initial state and the message buffer is empty.
	
	- \textit{Resulting}. A configuration that is a result from execution of any schedule, regardless of the number of events.
	
	- \textit{Reachable}. A configuration is reachable if it is a result from execution of a schedule with a finite number of events.
	
	- \textit{Accessible}. A configuration is accessible if it is reachable from some initial configuration.
	
	- \textit{Decision}. An accessible configuration where at least $ (N-1) $ processes have reached decision state.
	
	Run types:
	
	- \textit{Admissible run}. A run starting from an initial configuration where at most one process is faulty and all messages sent to correct processes are received.
	
	- \textit{Deciding run}. An admissible run that has reached a decision configuration. 
	
	Correctness Condition:
	
	- \textit{Totally correct protocol}. A consensus protocol is totally correct if every admissible run is a deciding run with one decision state.	
	
	\textcolor{MyBlue}{\section{VECTOR CONSENSUS ALGORITHM}}
	
		This algorithm implements consensus with vector agreement, as per Definition 3, in the defined model. It ensures that every admissible run is a deciding run (liveness) and all correct processes reach the same decision state (safety). The algorithm operates in three phases: Initial, Proposals, and Decision. Everything specific for it happens in the Proposals and Decision phases but it cannot operate without the Initial phase. 
	
		The algorithm implements \textit{reliable broadcast}. On receiving a message directly from its originator, every process broadcasts it to all processes, except to the originator. From start to termination, a process originates 4 messages: \textit{initial value}, \textit{first proposal}, \textit{second proposal}, and \textit{decision seed}. Handling of the received messages is regulated by 6 rules: Order Rule, Blend Rules 1 and 2, Completion Rules 1 and 2, and Update Rule. 
	
	\textcolor{MyBlue}{\subsection{\textsf{Algorithm Phases}}}
	
		\textit{Clarification}: A process does not receive messages from itself.
	
	
		\textcolor{MyBlue}{\textbf{Initial Phase}}
	
		Every process distributes its initial input value to the rest of the system's processes. A process $P_i$ starts it by broadcasting an \textit{initial value} message. In a system of \textit{N} processes, $P_i$ completes it on receiving \textit{initial value} messages from $(N-2)$ processes.
	
		\textcolor{MyBlue}{\textbf{Proposals Phase}}
	
		The processes distribute and start handling previously and currently \textit{first proposal} and \textit{second proposal} messages. Process $P_i$ starts it by broadcasting a \textit{first proposal} and completes it either with a vector $V_i$ = $V^F$, i.e. having no $\varnothing$ element, or with $V_i$ = $V^{\varnothing}$. Vector $V^{\varnothing}$ can have $N$ possible values, each with a $\varnothing$ element in a different position. This phase narrows the set of possible $V_i$ values from $(N+1)$ to just 2, i.e., one value of $V^{\varnothing}$ and $V^F$. $P_i$ completes it on receiving $(N-2)$ equal \textit{first proposal} or \textit{second proposal} messages from $(N-2)$ processes.

		\textcolor{MyBlue}{\textbf{Decision Phase}}
	
		The processes distribute, receive, and handle \textit{decision seed} messages to decide on the value of decision vector $V^{De}$. $P_i$ starts it by broadcasting a \textit{decision seed} message containing $V_i$. Some processes $P_j$ start it with $V_j$ = $V^{\varnothing}$, other processes $P_k$ start it with $V_k$ = $V^F$. Yet all correct processes complete this phase with the same decision vector $V^{De}$ and terminate the consensus round.
	
	\textcolor{MyBlue}{\subsection{\textsf{Algorithm Messages}}}
	
		\textcolor{MyBlue}{\textbf{Initial Value}}
		
		An \textit{initial value} message is broadcast by a process $P_i$ at the start of the Initial Phase to distribute its initial input value $v_i$, loaded with the content of $R_i^{In}$, to the rest of processes in the consensus system. An \textit{initial value} message received by $P_i$ after it has completed the Initial Phase is recorded in memory but ignored.
	
		\textcolor{MyBlue}{\textbf{First Proposal}}
	
		A process $P_i$ broadcasts a \textit{first proposal} message, $\Pi_i^{Frst}$, to start the Proposals Phase. $\Pi_i^{Frst}$ message is loaded with $V_i$ = $V^{\varnothing}$, which contains the $(N-1)$ initial input values known by $P_i$ at the end of Initial Phase, on their respective positions as vector elements. $\Pi_i^{Frst}$ \textbf{always contains} a $ \varnothing $ vector element.
	
		\textcolor{MyBlue}{\textbf{Second Proposal}}
	
		After a process $P_i$ has broadcast $\Pi_i^{Frst}$, it may broadcast a \textit{second proposal} message, $\Pi_i^{Scnd}$ obeying the Blend Rule 1 or the Blend Rule 2, described in Subsection 4.3. Process $P_i$ may broadcast $\Pi_i^{Scnd}$ before or after completion of Proposals Phase. $\Pi_i^{Scnd}$ is always loaded with $V_i$ = $V^{F}$, i.e., it \textbf{never contains} a $ \varnothing $ vector element.
	
		\textcolor{MyBlue}{\textbf{Decision Seed}}
	
		On completion of the Proposals Phase, $P_i$ starts the Decision Phase by broadcasting a \textit{decision seed} message $\Delta_i$. It contains the vector value $V_i$. The broadcast $\Delta_i$ and the received $\Delta$ messages allows $P_i$ to confidently decide $V^{De}_i$ that is the same as the $V^{De}$ decided by the rest of the correct processes.
	
	\textcolor{MyBlue}{\subsection{\textsf{Algorithm Rules}}}
	
	\textcolor{MyBlue}{\textbf{Order Rule}}
	
	Order Rule: The handling order of the received messages is in the order of sending and not before the phase related to them.	
	
	\textit{Effect}: It neutralizes two effects of asynchrony: i) processes do not start a round of the algorithm simultaneously; and ii) messages may be received not in the order of sending. 
	
	\textit{Example}: While in the Initial Phase, process $P_j$ receives the \textit{second proposal} message of process $P_i$ and later the \textit{first proposal} message of $P_i$. Process $P_j$ records both messages in the memory and handles them after entering Proposals Phase. $P_j$ handles the \textit{first proposal} message first and the \textit{second proposal} message second. 
	
	\textcolor{MyBlue}{\textbf{Blend Rules}}
	
	\textit{Effect}: Enforce indirect distribution of an \textit{initial value}, directly received (Blend Rule 1) or indirectly received (Blend Rule 2) from the original broadcaster of the \textit{initial value} message. 
	
	\textit{Note 1}: In contrast, \textit{reliable broadcast} sends messages indirectly and thereby pursues different objectives.
	
	\textit{Note 2}: For equal influence of directly and indirectly received messages, applying any of both Blend Rules requires a process to have received $(N-2)$ retransmitted \textit{first proposal} messages from each one of $(N-2)$ processes.
	
	\textcolor{MyBlue}{\textbf{Blend Rule 1}}: On handling a received \textit{first proposal} that is different from the recipient's \textit{first proposal}, the recipient that has not yet broadcast a \textit{second proposal} message must do it.
	
	\textit{Example}: Process $P_i$, having received from a process $P_j$ a $\Pi_j^{Frst}$ that contains a $ \varnothing $ vector element on a position with a different index compared to the position index of the $ \varnothing $ vector element in $\Pi_i^{Frst}$, prepares and broadcasts a $\Pi_i^{Scnd}$ message, even after $P_i$ has completed Proposals Phase.
	
	\textcolor{MyBlue}{\textbf{Blend Rule 2}}: On handling a \textit{second proposal} message, a process that has not yet broadcast a \textit{second proposal} message must do it.
	
	\textit{Example}: Process $P_i$ that has not yet broadcast a \textit{second proposal}, having handled a $\Pi_j^{Scnd}$ message from $P_j$, prepares and broadcasts a $\Pi_i^{Scnd}$ message, even after $P_i$ has completed Proposals Phase.
	
	\textcolor{MyBlue}{\textbf{Completion Rule 1}}
	
	Completion Rule 1: On receiving $(N-2)$ equal \textit{first proposal} messages, the receiver must complete Proposals Phase with $V^\varnothing$.
	
	\textit{Effect}: Enforces completion of Proposals Phase with $V^\varnothing$. 
	
	\textit{Example}: Process $P_i$ completes Proposals Phase with a vector value $V_i$ = $V^\varnothing$ after it has received $(N-2)$ equal $\Pi^{Frst}$ messages.
	
	
	\textcolor{MyBlue}{\textbf{Completion Rule 2}}
	
	Completion Rule 2: On receiving $(N-2)$ \textit{second proposal} messages, the receiver must complete Proposals Phase with $V^F$.
	
	\textit{Effect}: Enforces completion of Proposals Phase with $V^F$. 
	
	\textit{Example}: Process $P_i$ completes Proposals Phase with a vector value $V_i$ = $V^F$ after it has handled $(N-2)$ messages $\Pi^{Scnd}$.
	
	\vspace{\baselineskip}
	
	\textcolor{MyBlue}{\textbf{Update Rule}}
	
	Update Rule: A process starting Decision Phase with $V^{\varnothing}$ must update it to $V^F$ on receiving a $\Delta$ message containing $V^F$. 
	
	\textit{Effect}: Ensures that all correct processes complete this phase and terminate the algorithm with $V^{De}=V^F$ if $\ge 2$ processes started Decision Phase with $V^F$; otherwise terminate with $V^{De}=V^{\varnothing}$.
	
	\textit{Example}: Process $P_i$ completes Decision Phase with a vector value $V^F$ if it started the phase with $V^F$ or if the Update Rule was applied. Otherwise, $P_i$ completes the phase with a vector value $V^{\varnothing}$.
	
	
	\textcolor{MyBlue}{\subsection{\textsf{Algorithm Operation}}}
	
		Algorithm messages are received directly from their originators and indirectly via retransmission by the direct recipients. On receiving a message directly, the recipient multicasts it to all other processes except the originator. The content of the messages related to previous consensus rounds is ignored. The content of messages related to the current or next rounds is recorded and processed according to the Order Rule.
	
		\textcolor{MyBlue}{\textbf{Initial Phase}}
	
		A process $P_i$ starts it by broadcasting an \textit{initial value} message. In a system of $N$ processes, $P_i$ completes this phase on receiving \textit{initial value} messages from $(N-2)$ processes and enters Proposals Phase. 
	
		\textcolor{MyBlue}{\textbf{Proposals Phase}}
	
		$P_i$ starts it by broadcasting a \textit{first proposal} message and obeying Order Rule, handles the \textit{first proposal} and \textit{second proposal} messages received before, during, or after this phase. Obeying Order Rule and: i) either Blend Rule 1 or Blend Rule 2, $P_i$ broadcasts a \textit{second proposal} message during this phase or after it completes; ii) either Completion Rule 1 or Completion Rule 2, on receiving $(N-2)$ equal \textit{first proposal} or \textit{second proposal} messages, $P_i$ completes this phase with either $V_i$ = $V^\varnothing$ or $V_i$ = $V^F$. Broadcasting \textit{second proposal} after completion ensures that all correct processes enter the Decision Phase.
	
		\textcolor{MyBlue}{\textbf{Decision Phase}}
	
		$P_i$ starts it by broadcasting a \textit{decision seed} message. It contains the vector value $V_i$. On receiving $(N-2)$ \textit{decision seed} messages $P_i$ confidently decides $V^{De}_i$ knowing that it is the same as the $V^{De}$ decided by the rest of the correct processes.
	
		\textcolor{MyBlue}{\textbf{Operational Guarantees}}
	
		The Proposals Phase guarantees the algorithm's liveness. It completes with prepared groundwork for safety. It ensures 3 critical certainties: i) every correct process completes it with either $V^\varnothing$ or $V^F$; ii) all processes, completing with $V^\varnothing$, do it with the same value of $V^\varnothing$; and iii) only one process completing with $V^F$ is impossible.
	
		The Decision Phase guarantees the algorithm's safety without risking its liveness. It ensures one operational certainty that is critical for safety: all correct processes in the system terminate with the same decision value $V^{De}$.

	\textcolor{MyBlue}{\section{PROOFS}}
	
	\textcolor{MyBlue}{\subsection{\textsf{Termination with Valid Vector Agreement}}} 
	
	\textbf{Theorem 1}: \textit{A vector consensus protocol can operate with ensured simultaneous safety and liveness in total asynchrony in spite of a possible process crash}.
	
	\begin{proof}
	
	Consensus requires every process to distribute to every other process the data in its $R^{In}$ and all correct processes agree about ordering of individual processes' data in a single dataset. Sufficient is to show the ability to terminate reaching a decision configuration with only one decision state. The proof follows from these 4 lemmas:
	
	\end{proof}
	
	\textbf{Lemma 1}: \textit{A protocol can ensure that every non-crashed process enters Decision Phase.}
	
	\begin{proof}	
	
	Entering Decision Phase requires completion of Proposals Phase. We will show that every correct process $P_i$ that cannot complete under Completion Rule 1 completes under Completion Rule 2. 
	
	Receiving $(n-2)$ equal $\Pi^{Frst}$ messages that originated from $(n-2)$ processes triggers Completion Rule 1. This cannot take place when less than $(n-2)$ processes $P_j, j \ne i,$ have originated and broadcast equal $\Pi^{Frst}$ before one of them eventually crashes. 
	
	Receiving $(n-2)$ equal $\Pi^{Scnd}$ messages that originated from $(n-2)$ processes triggers Completion Rule 2. Sufficient condition is $(n-2)$ processes $P_k, k \ne i,$ to have originated and broadcast $\Pi^{Scnd}$ messages before one of them eventually crashes. 
	
	Broadcasting a $\Pi^{Scnd}$ triggered by Blend Rule 1 requires at least one process to have received $\Pi^{Frst}$ not equal to its own $\Pi^{Frst}$. Sufficient condition for all non-crashed processes to broadcast a $\Pi^{Scnd}$ triggered by Blend Rule 2 is one process to have broadcast a $\Pi^{Scnd}$ triggered by Blend Rule 1.
	
	If Completion Rule 1 cannot be satisfied, the possible cases are: i) A set of $(n-2)$ processes $P_j$ originated and broadcast equal $\Pi_j^{Frst}$ and $P_i$ is in this set; or ii) A set of $(n-3)$ or less processes $P_j$ originate and broadcast equal $\Pi^{Frst}$ and $P_i$ is not in this set.
	
	In case (i), two processes have originated and broadcast messages $\Pi^{Frst} \ne \Pi_i^{Frst}$. In case (ii), two or more processes have originated and broadcast messages $\Pi^{Frst} \ne \Pi_i^{Frst}$. In both cases, at least $P_i$ originates and broadcasts $\Pi_i^{Scnd}$ triggered by Blend Rule 1. As a consequence, on receiving $\Pi_i^{Scnd}$, no less than $(n-2)$ processes originate and broadcast $\Pi^{Scnd}$ triggered by Blend Rule 2.
	
	Thus, Blend Rule 1 and Blend Rule 2 ensure that $P_i$ completes Proposals Phase. If $P_i$ cannot complete under Completion Rule 1, it completes under Completion Rule 2.
	
	\end{proof}
	
	\textbf{Lemma 2}: \textit{A protocol can ensure that every correct process starts deciding either with all initial inputs or lacking the same initial input.}
	
	\begin{proof}
	
	A process completes Proposals Phase with a $V^\varnothing$ vector when it has received and handled ($N-2)$ equal $\Pi^{Frst}$ messages, i.e., each containing a $V^\varnothing$ vector having a $\varnothing$ element with the same index. In a system of $N$ processes, where $N \ge 5$, this itself precludes the possibility for existence of another ($N-2)$ processes that have broadcast equal $\Pi^{Frst}$ messages with a different $V^\varnothing$ vector.
	
	\end{proof}
	
	\textbf{Lemma 3}: \textit{A protocol can ensure that either more than one process or none starts deciding with a vector having all initial inputs.}
	
	\begin{proof}
	
	Let $E$ be a set of processes that prepare equal $\Pi^{Frst}$ messages. Assume existence of configuration where only process $P_i$ enters Decision Phase with vector $V^F$ and the rest of processes with $V^\varnothing$. 
	
	\textit{Cases With No Crash}
	
	With no crash, the assumption requires $(N-1)$ processes in $E$. Otherwise, $(N-1)$ processes cannot enter with $V^\varnothing$.
	
	Let $P_i \notin E$. Process $P_i$ being the only one entering Decision Phase with vector $V^F$ means that every process $P_j \in E$ completes Proposals Phase with vector $V^\varnothing$ under Completion Rule 1. The necessary condition is that every $P_j$ has received $(N-2)$ equal $\Pi^{Frst}$ messages. In this case $P_i$ also receives $(N-2)$ equal $\Pi^{Frst}$ messages and enters Decision Phase with vector $V^\varnothing$. 
	
	Let $P_i \in E$ and the broadcast by process $P_s \in E$ message $\Pi_s^{Frst}$ is delayed, so that no process in $E$ receives $(N-2)$ equal $\Pi^{Frst}$ messages, except $P_s$ and also the process $P_k \notin E$. Processes $P_s$ and $P_k$ complete under Completion Rule 1 with vector $V^\varnothing$. Process $P_i$ receives from $P_k$ message $\Pi_k^{Frst} \ne \Pi_i^{Frst}$. Triggered by Blend Rule 1, $P_i$ prepares and broadcasts $\Pi_i^{Scnd}$. On receiving $\Pi_i^{Scnd}$, every process prepares and broadcasts a $\Pi_i^{Scnd}$ triggered by Blend Rule 2. Thus the $(N-2)$ processes in $E$ that are still in Proposals Phase complete it with $V^F$ on receiving $(N-2)$ messages $\Pi^{Scnd}$.
	
	Let $P_i \in E$ and only $P_i$ does not receive $\Pi_s^{Frst}$ directly from $P_s$ before receiving $(N-2)$ $\Pi^{Frst}$ messages. $P_i$ will receive $\Pi_s^{Frst}$ indirectly retransmitted before Blend Rule 1 is allowed to trigger preparing and broadcasting $\Pi_i^{Scnd}$ in response to received $\Pi_k^{Frst}$. Thus $P_i$ also receives $(N-2)$ equal $\Pi^{Frst}$ messages and also enters the Decision Phase with vector $V^\varnothing$. 
	
	\textit{Cases With A Crash}
	
	Let a process $P_c$ crashes before broadcasting it \textit{initial value}. So, all $(N-1)$ correct processes are in $E$ and broadcast $(N-1)$ equal $Pi^{Frst}$ messages. Thus all $(N-1)$ processes individually receive equal $Pi^{Frst}$ messages, complete Proposals Phase under Completion Rule 1, and enter Decision Phase with vector $V^\varnothing$.
	
	Let $P_i \in E$, $(N-1)$ processes are in $E$, and a process $P_c \in E$ crashes \textbf{before} broadcasting its $\Pi_c^{Frst}$. The rest of processes in $E$, including $P_i$, receive $(N-2)$ equal $\Pi^{Frst}$ messages and enter Decision Phase with vector $V^\varnothing$. Process $P_k \notin E$ also receives $(N-2)$ equal $\Pi^{Frst}$ messages and enters Decision Phase with vector $V^\varnothing$.
	
	Let $P_i \in E$, $(N-1)$ processes are in $E$, and a process $P_c \notin E$ crashes \textbf{after} broadcasting its $\Pi_c^{Frst}$. All processes in $E$ receive $(N-2)$ equal $\Pi^{Frst}$ messages, complete under Completion Rule 1, and enter Decision Phase with $V^\varnothing$. 
	
	Let $P_i \notin E$, $(N-1)$ processes are in $E$, and a process $P_c \in E$ crashes \textbf{before} broadcasting its $\Pi_c^{Frst}$. $P_i$ receives $(N-2)$ equal $\Pi^{Frst}$ messages, completes under Completion Rule 1, and enters Decision Phase with vector $V^\varnothing$. On receiving $\Pi_i^{Frst}$ originated by $P_i$, the processes in $E$, individually trigger Blend Rule 1, enforcing each one to prepare and broadcast $\Pi^{Scnd}$. On receiving a $\Pi^{Scnd}$, process $P_i$ triggers Blend Rule 2, which enforces it to prepare and broadcast $\Pi_i^{Scnd}$. Thus every correct process in $E$ receives $N-2$ $\Pi^{Scnd}$ and completes under Completion Rule 2 with a vector $V^F$.
	
	Let $P_i \notin E$, $(N-2)$ processes are in $E$, and a process $P_c \notin E$ crashes \textbf{after} broadcasting its $\Pi_c^{Frst}$. Process $P_i$ receives $(N-2)$ equal $\Pi^{Frst}$ messages, completes under Completion Rule 1, and enters Decision Phase with $V^\varnothing$. On receiving $\Pi_i^{Frst}$ originated by $P_i$, the processes in $E$, individually trigger Blend Rule 1, enforcing each one to prepare and broadcast $\Pi^{Scnd}$. On receiving a $\Pi^{Scnd}$, process $P_i$ triggers Blend Rule 2, which enforces it to prepare and broadcast $\Pi_i^{Scnd}$. Thus all $(N-2)$ processes in $E$ receive $(N-2)$ $\Pi^{Scnd}$ and complete under Completion Rule 2 with vector $V^F$.
	
	\textit{Recapitulation}
	
	In all relevant cases with no crash, the possibilities are two: i) either $N$ processes complete Proposals Phase with vector $V^\varnothing$; or ii) $(N-2)$ processes complete it with vector $V^F$ and 2 with vector $V^\varnothing$. In all relevant cases with one crash, the possibilities are also two: i) either $(N-1)$ processes complete Proposals Phase with $V^\varnothing$; or ii) $(N-2)$ processes complete it with vector $V^F$ and 1 with vector $V^\varnothing$. Hence the assumption is incorrect. 
	
	\end{proof}
	
	\textbf{Lemma 4}: \textit{A protocol can terminate with all correct processes agreed either on all initial inputs or on lacking the same initial input.}
	
	\begin{proof}
	
	Decision Phase may have the following configurations:
	
	- \textit{Case A}: Only one process started with $V^F$. This possibility is precluded [Lemma 3].
	
	- \textit{Case B}: $\ge 2 $ processes started with different  $V^{\varnothing}$ vectors. This possibility is precluded [Lemma 2].
	
	- \textit{Case C}: $\ge 2 $ processes started with $V^F$. All correct processes broadcast a $\Delta$ message. Each one that started with $V^{\varnothing}$ is guaranteed to receive one or more message containing a $V^F$ vector and, obeying the Update Rule, updates its vector from $V^{\varnothing}$ to $V^F$. Thus every correct process, knowing that every other correct process either started with $V^F$ or has updated to $V^F$, terminates with $V^{De}$ = $V^F$.
	
	- \textit{Case D}: All processes started with $V^{\varnothing}$. All correct processes broadcast a $\Delta$ message and no correct process receives a message with vector $V^F$. Thus, knowing that Decision Phase cannot start with less than two processes with vector $V^F$, every correct process knows that all correct processes have started with a $V^{\varnothing}$ and terminates with $V^{De}$ = $V^{\varnothing}$.	
	
	\end{proof}
	
	\textcolor{MyBlue}{\subsection{\textsf{Contrast with Crusader Agreement}}}
	
	In the context of a system with one possible crash-fail, Crusader agreement \cite{Dolev1981} implies the possibility for existence of two sets of processes, formed in the following manner as a result of the crash. A process successfully sends its message to a fraction of all processes immediately before crashing. All processes in this fraction form the first set. All processes outside it form the second set. Crusader agreement allows the processes of the second set to enter a decision state with $ \varnothing $ in place of the missing initial value, which the processes were supposed to receive from the crash-failed process. This might create a false sense of similarity with our solution.
	
	The correctness condition of our system model does not allow a weakened agreement property, i.e., a consensus protocol is correct only if every admissible run is a deciding run with one decision state – the same for all processes. As Theorem 1 demonstrates, the presented algorithm ensures that all correct processes, in cases with or without a crash, decide exactly the same agreement value. With no crash, the input value of a slow process, or one with slow output links, can be excluded from the agreement value. Yet, in spite of that, this process unconditionally enters a decision state with the same agreement value as the rest of the processes.
	
	\vspace{\baselineskip}

	\textcolor{MyBlue}{\subsection{\textsf{Termination with Binary Agreement}}} 
	
	Here we show that every leaderless consensus, and in particular one with binary agreement, reduces to vector consensus. We show that, a crash-tolerant in total asynchrony, consensus protocol with binary agreement ensures entering a decision configuration with one decision state by using our crash-tolerant algorithm with vector agreement.The proof relies on the concept of atomic consistency across multiple copies of the same data and on the concept of state machine replication. So, they need to be presented briefly.
	
	Atomic (a.k.a. linearisable) consistency \cite{Herlihy_Wing_1990} of data replicated on multiple computers is nearly what the sequential consistency \cite{Lamport_1979} is for a multiprocessor computer that concurrently executes multiple programs with update-access over the same data. The difference: atomic consistency requires every replica to implement sequential consistency under exactly \textit{the same} sequential order as the rest.
	
	State machine replication \cite{Schneider_1990} \cite{Lampson_1996} is the known way to arrange replicas, built as deterministic state machines \cite{Lamport_StateMachine}, to do exactly the same thing. Thus replicas starting from exactly the same state and doing exactly the same thing finish with exactly the same state. Hence, replicas having started from an atomically consistent state and having transformed it with a deterministic rule finish with an atomically consistent state.
	
	\vspace{\baselineskip}
	
	\textbf{Theorem 2}: \textit{A consensus protocol with binary agreement can be totally correct in total asynchrony with possible crash of one process.}
	
	\begin{proof}
	
	The idea is to show that protocol's every admissible run is a deciding run with only one decision state. The proof follows from the following two lemmas:
	
	\end{proof}
	
	\textbf{Lemma 5}: \textit{Processes of a totally correct consensus protocol with binary agreement decide from atomically consistent states.}
	
	\begin{proof}
	
	Assume not. By definition, a consensus protocol is totally correct if every admissible run is a deciding run with only one decision state. A deciding run is an admissible run that has reached a decision configuration. A decision configuration is a configuration where all non-crashed processes have reached a decision state.
	
	The correctness of a protocol requires all correct processes to reach an atomically consistent binary outcome. This might happen by chance, i.e., with probability smaller than 1. In this case not every admissible run is a deciding run with only one decision state.
	
	Atomically consistent binary outcome with every deciding run requires: i) all correct processes to have produced atomically consistent individual datasets; and ii) every process to produce its binary outcome by passing its dataset to its own instance of the same deterministic procedure. Hence the assumption is incorrect.
	
	\end{proof}
	
	\textbf{Lemma 6}: \textit{A binary consensus protocol, using a fault-tolerant in asynchrony consensus algorithm with ensured safety and liveness, operates with ensured safety and liveness.}
	
	\begin{proof}
	
	A consensus algorithm with vector agreement and tolerance to one process crash in total asynchrony can ensure termination with atomically consistent vector values [Theorem 1].
	
	Termination with atomically consistent vector values guarantees termination with atomically consistent binary values [Lemma 5]. protocol that ensures termination with atomically consistent decision values is totally correct by definition.
	
	\end{proof}

	\textcolor{MyBlue}{\section{PHYSICAL EXPERIMENTS}}
	
	We performed physical experiments to demonstrate with experimentally obtained data that impossibility of binary consensus is not a proof of impossibility of vector consensus. The same \textit{bivalent} configuration presented in the FLP work, was also demonstrated in a fully synchronous system in the work of Santoro and Widmayer (SW) \cite{SantoroEtAl1989}. Both systems are equivalent in the sense of implementing each other \cite{Gafni_Losa_2023}, but the synchronous one is easier to experiment with and reason about.
	
	SW work studies agreement in synchrony in presence of \textit{dynamic} transmission faults. It demonstrates that the additional destructive capability that these faults inherit from being dynamic overwhelms any additional power the system may derive from complete synchrony. A system of $N$ processes can model dynamic transmission faults in the following manner: i) starts a messaging round with a set of $(N-1)$ faulty links \cite{SantoroEtAl1989}; and ii) switches to a different set of $(N-1)$ faulty links before the next round.
	
	\textcolor{MyBlue}{\subsection{\textsf{The Experiments}}}	
	
	We performed the following set of experiments with a fully synchronous system of $5$ processes – attempted to reach binary agreement individually with each one of all combinations with the critical minimum of $4$ faulty links that prevent it, out of the $20$ total links, i.e., with all combination of 16 correct links. The total number of combinations $20C16$ is $20! / (16! * (20 - 16)!)$, i.e., 4,845.
	The total number of performed experiments required to explore $20C16 * 20C16$ combinations, i.e., $23,474,025$. On a computer with 24 hardware threads, the exploration of this number of combinations lasted around 5 hours.
		
	Each experiment was performed in two steps. During the First step the processes broadcast their initial input values with one combination of faulty links. During the Second step every process broadcasts all initial input values it knows at the end of First step. The Second step executes two more messaging rounds necessary for termination of consensus protocol. Thus, one experiment comprises one execution of First step and 4,845 executions of Second step. The entire set comprises 4,845 performed experiments. 
	
	We performed the experiments with two different algorithms. Each one required 3 processes to agree on the same binary value. One explored for consensus terminated with binary agreement. The other explored for consensus terminated with vector agreement and computed binary value from the agreed vector.
	
	\textcolor{MyBlue}{\subsection{\textsf{Possibility for Scaling Up?}}}
	
	Performing exactly the same set of experiments, with a system of 7 processes is not a realistic possibility. It requires exploration of $42C36 * 42C36 = 5,245,786 * 5,245,786 = 27,518,270,757,796$ combinations, which is $1,172,286$ times more computations than with a system of 5 processes. This would require $1,172,286 * 5 = 5,861,430$ hours to complete. Thus experimentation with a system of 7 processes would last 669 years.
	
	Performing these experiments, with a system of 6 processes implemented with the same software is more realistic, It requires exploration of $30C25 * 30C25 = 20,307,960,036$ combinations and needs 805 times longer time than a system of 5 processes needs to complete, i.e., $805 * 5 = 4025$ hours. So, experimentation with a system of 6 processes would last 5 months and 2 weeks uninterrupted computation. 
		
	In the context of binary agreement under the FLP or SW models, the outcomes of eventual experimentation with a system of 6 processes became to a certain degree predictable after the completion of experiments with a system of 5 processes. Even when every process receives the initial inputs of all processes in the system, binary agreement will be frequently impossible due to the inevitably high probability for \textit{tie state}. Thus, the experimental results will not reveal anything unknown in advance.
	
	
	\textcolor{MyBlue}{\subsection{\textsf{The Results}}}	
	
	The results demonstrate that binary consensus terminated with \textit{binary} agreement is impossible when one process cannot decide due to \textit{tie score}, two processes vote for 0 and two processes vote for 1. In contrast, binary consensus terminated with \textit{vector} agreement in theory becomes impossible in two cases: i) vector agreement is impossible; and ii) vector agreement is possible, but the agreed vector contains 4 input values, two of them 0 and two of them 1. Yet no experimental configuration prevented binary agreement by preventing the vector agreement.

	\begin{table}[ht]
		\caption{Vector and Binary Agreements in Total Synchrony}
		\begin{tabular}{l r}
			\hline 
			\\
			Combinations with 4 Faulty Links = 20C16			&      4,845 \\
			Configurations Explored = 20C16 x 20C16				& 23,474,025 \\
			\textbf{Vector Agreement}:							&			 \\
			Configurations with Agreement						& 23,474,025 \\
			\textbf{Binary Agreement}:							&			 \\
			From \textbf{5} Initial Inputs 
			with \textbf{5} Agreed Processes					& 23,473,682 \\
			From \textbf{4} Initial Inputs
			with \textbf{5} Agreed Processes					&        134 \\
			From \textbf{4} Initial Inputs
			with \textbf{No} Agreement							&        209 \\
			\\
			\hline
		\end{tabular}		
	\end{table} 
	
	The experimental results presented in Table 1 demonstrate that in a completely synchronous system of $5$ processes, the existence of set of $4$ \textit{dynamic} faulty links:
	
	1. Cannot make vector agreement impossible.
	
	2. Can make binary agreement impossible when the binary value is computed from an even number of initial inputs.
	
	The number of configurations with no agreement is different with the algorithm terminating with binary agreement. Analysis of the differences is beyond the objectives of this paper. Important is the confirmation that with both algorithms the impossibility of binary consensus is caused entirely by  the interpretation of \textit{tie state} as "undecidable".
	
	\textcolor{MyBlue}{\section{DISCUSSION}}
	
	
	\textcolor{MyBlue}{\subsection{\textsf{Analysis of the Experimental Results}}}
	
	The results demonstrate that nothing prevents vector consensus under the configurations making binary agreement unreachable and let us clarify what causes the impossibility result. Proving the existence of one-way causal relationship \cite{Menzies2001CounterfactualTO} between cause $(C)$ and effect $(E)$ requires to demonstrate existence of two dependencies: i) when $C$ is present, $E$ is also present; and ii) when $C$ is not present, $E$ is also not present. 
	
	If impossibility of binary consensus were caused by asynchrony, then in absence of asynchrony (e.g., in complete synchrony) binary consensus must be possible. Since SW demonstrated impossibility in complete synchrony, it follows that asynchrony is not the cause. If impossibility of consensus, both is synchrony and asynchrony, were caused by data-dependence of agreement (e.g., with binary agreement), then the consensus must be impossible with binary agreement in synchrony and asynchrony, and must be possible with data-independent agreement (e.g., with vector agreement).
	
	FLP and SW demonstrated the impossibility of binary consensus in synchrony and asynchrony. We demonstrated, analytically in asynchrony and experimentally in synchrony, that vector consensus is possible. From these results it follows that the FLP impossibility is caused by the data-dependence of binary agreement rather than by asynchrony. It also follows that evidence against the possibility of vector, or any other data-independent, consensus does not exist.
	
	\textcolor{MyBlue}{\subsection{\textsf{Discussion of FLP Proof}}}
	
	The proof of FLP theorem presents the following evidence:
	
	1. A configuration with 3 possible outcomes – 0, 1, or \textit{tie state}  – inevitably happens and was given the inappropriate name \textit{bivalent} configuration, even though the configuration is in fact \textit{trivalent}.
	
	2. \textit{Tie state} is considered  "undecidable" despite the necessity of a fault-tolerant binary consensus protocol to be able to terminate with this outcome if one of an odd number of processes crashes or none of an even number of processes does.
	
	3. The FLP system model assumes \textit{atomic broadcast}. Under this assumption, if one process has received the initial input from the slow process, all processes will receive it. If none has, all processes can safely terminate with no input from the slow process, including the slow process itself. This handles also the case with a crash. The FLP proof ignores this possibility.
	
	4. The processes in the proof do not attempt to discover what set of messages the other processes have received and thus establish consensus with initial inputs from all processes and handle the case where the slow process is not the same for all processes.
	
	5. The proof implicitly suggests, without any evidence, that the shortcut declaring impossibility of consensus is the only existing option, instead of coordinating the decision making with the other processes. 
	
	The proof \textbf{does not} present any evidence that: 
	
	1. The \textit{trivalent} configuration is impossible to manage in total asynchrony. 
	
	2. Fault-tolerance is possible without recognition of \textit{tie state} as legitimate outcome for termination purposes, before using it as indication of impossibility of binary agreement. 
	
	3. An eventual impossibility of \textit{binary} consensus indicates, reveals, causes, or proves impossibility of \textit{any} consensus.
	
	\textcolor{MyBlue}{\subsection{\textsf{Reconciliation}}}
	
	In a system of $n$ processes, a fault-tolerant consensus algorithm with vector agreement about the inputs from the system processes must be able to terminate with agreement about the inputs of $(n-1)$ processes; otherwise the algorithm is not fault-tolerant. Proving that vector agreement is impossible in asynchrony requires to demonstrate that $(n-1)$ processes: i) cannot agree on a set of at least their own initial input values; or ii) cannot make sure that a process, which is not crashed but is slow or has slow outgoing links, decides and agrees on the same set of initial input values.
	
	The proof of the FLP theorem does not demonstrate impossibility of agreement about the initial inputs. It demonstrates only the inevitable existence of a trivalent configuration. This is not proof of impossibility of vector agreement. Table 1 shows that the existence of such configuration has no whatsoever effect on the possibility of vector agreement. Hence, there is no contradiction between the termination of our algorithm and what the FLP theorem in fact demonstrates as there is no intersection. The contradiction is between the shown termination with vector agreement and the claim of FLP theorem that it demonstrates impossibility to terminate with \textit{any} agreement.
	
	Kurt Gödel shows in his \textit{incompleteness theorems} \cite{Hofstadter_1979} that logical contradiction between two systems does not necessarily indicate that one of them is incorrect, but rather difference in the assumptions. Resolving the incompatibility needs a more flexible frame that includes the traditional principles, but also adds new dimensions.
	
	We introduced a new frame. It acknowledges the existence of two different consensus types: data-dependent and data-independent, and recognises that vector agreement is the most basic form of agreement. It focuses on how the FLP proof comes to the conclusion of impossibility and reveals its reliance on three unstated assumptions that are the reconciliation borderline. In absence of any of these assumptions the FLP impossibility result does not hold.
	
	Establishing the implicit assumptions behind the FLP impossibility result – the actual borderline separating the demonstrated possibility from the impossibility shown with the FLP theorem – is no less important for the theory than the demonstration of the possibility itself. This borderline allows the proven possibility and the FLP impossibility to coexist in peace in spite of the logical contradiction – each one being correct within its individual set of assumptions.
	
	\textcolor{MyBlue}{\subsection{\textsf{FLP Impossibility Is Model-Dependent, Not Absolute}}}

	The FLP impossibility result is model-dependent. Its model implicitly makes a choice to interpret certain configurations of the system state as “undecidable” and therefore preventing agreement. In the model, termination is defined as: "Every non-faulty process eventually decides a value", and decision is defined as: "Irrevocable choice of a value from $\{0,1\}$". Thus a coincidence of circumstances may create state configurations where agreement is made impossible, which in turn, under another implicit choice, makes termination logically impossible.
	
	\vspace{\baselineskip}
	\vspace{\baselineskip}
	
	\textcolor{MyBlue}{\textbf{\textsf{The Choice Making Agreement Impossible}}}

	This choice is about the interpretation of system state’s \textit{tie} configurations. \textit{Tie} is a fully legitimate state configuration of systems with an odd number of processes when one process crashes. Yet the model does not allow the \textit{tie} state to have a legitimate (binary) value assigned to it. Instead, the model chooses to interpret \textit{tie} configurations as “undecidable”. Thus, the impossibility does not follow solely from asynchrony and crash failures; it also requires interpretation of \textit{tie state} as impossibility to decide. 

	This choice is a hidden assumption. It reveals that the impossibility of agreement is not inherently a property of fault tolerance. It is a property of the model-made specific interpretation. Changing this interpretation makes binary agreement possible.

	This choice is provably wrong. It renders the model inappropriate for the purposes of fault-tolerant computing. It can be used to prove impossibility of consensus in absence of any unannounced faults in synchrony. After a crash becomes known to the healthy processes, systems of odd number of processes continue operating as systems of even number of healthy processes. In systems with an even number of processes, states with \textit{tie} configuration naturally exist without an unannounced faulty process. Hence, the formally correct under this choice impossibility of binary agreement is an artificial construct detached from reality.

	Thus the impossibility of agreement is not about asynchrony or faults; it is about how \textit{tie} is interpreted and entirely depends on this choice. When a model chooses to assign a legitimate value to \textit{tie} states, the impossibility of agreement does not exist.
	
	\vspace{\baselineskip}

	\textcolor{MyBlue}{\textbf{\textsf{The Choice Making Termination Impossible}}}
	
	The definition of termination in the model – “every non-faulty process eventually decides a value” – does not specify the type of value that has to be decided. Yet the model implicitly interprets “decides a value” as “decides a binary value”, which is another hidden assumption. It excludes the option to terminate with decision on a vector and compute binary decision value from the decided vector after termination.
		
	Thereby the FLP model prepares conditions for a logically following chain reaction. In asynchrony, impossibility of agreement makes termination inevitably impossible. The formal logic behind the impossibility to terminate is impeccable. The informal one is not. It enables a model-distorted reflection of the nature-established limits.
	
	\vspace{\baselineskip}

	\textcolor{MyBlue}{\textbf{\textsf{Final Remarks}}}

	The FLP theorem proves impossibility to terminate binary consensus with help of two implicit assumptions of its model. The proof is formally correct, despite its inadequate reflection of reality. Yet the theorem claims more than that. It asserts that the proof demonstrates impossibility of \textit{any} fault-tolerant consensus in total asynchrony – a hypothesis with no evidence of support. Essentially, the theorem claims are two:

	- Claim 1: Impossibility with \textit{binary} agreement. Under the implicit assumptions of its model, the theorem proves it with strict correctness. This impossibility, however, is not universal. It is \textbf{model-dependent}, hence \textbf{not absolute}.

	- Claim 2: Impossibility with \textit{any} agreement. It does not follow logically from the correctness of Claim 1. It is an \textbf{overstatement} of what the theorem proves. Causality exists in the opposite direction. Impossibility with data-independent (e.g., vector) agreement causes impossibility with data-dependent (e.g., binary) agreement.
		
	In contrast, the demonstrated in the paper possibility of binary consensus requires assignment of binary value to \textit{tie state}. It follows from the demonstrated possibility of vector consensus that is \textbf{not model-dependent}, hence \textbf{absolute}.
	
	
	\textcolor{MyBlue}{\subsection{\textsf{Conclusion}}}
	
	The presented work reopened the studying of deterministic fault-tolerant consensus problem in total asynchrony. Until now the topic was considered settled by the famed FLP impossibility result, once and forever. We have shown that the strictly proven impossibility is model-dependent and not reflecting nature-established limits.
	
	This discovery has a major theoretical consequence. The perceived unconditional impossibility of a deterministic consensus protocol to operate with simultaneously ensured safety, liveness, and fault-tolerance in asynchrony lost its theoretical base. This consequence is significant for a variety of industries. It demonstrates the conceptual possibility to provide the same operational guarantees in the more important for practice partial synchrony.
	
	\textcolor{MyBlue}{\section*{Acknowledgement}}
	 
		The author expresses gratitude to Simeon Simoff for his help in smoothing out the paper's narrative and presentation.
	
	\textcolor{MyBlue}{\section*{Funding}}
	
		This work is partly sponsored by the Australian Federal Government through the Research and Development Tax Incentive Scheme. 
		
	\vspace{\baselineskip}
	\vspace{\baselineskip}
	
	\bibliographystyle{ACM-Reference-Format}
	\bibliography{CiA}

	\end{document}